\documentclass[12pt]{iopart}

\usepackage{iopams}  
\usepackage[dvips]{graphicx}
\usepackage{booktabs}
\usepackage{psfrag}

\begin{document}

\title{Interfacing light and single atoms with a lens}

\author{Meng Khoon Tey$^1$, Gleb Maslennikov$^1$, Timothy C. H. Liew$^1$, Syed Abdullah Aljunid$^1$, Florian Huber$^2$,
  Brenda Chng$^1$, Zilong Chen$^3$, Valerio Scarani$^{1,4}$ and Christian Kurtsiefer$^{1,4}$}

\address{$^1$ Center for Quantum Technologies, National University of Singapore,
3 Science Drive 2, Singapore, 117543}
\address{$^2$ Department of Physics, Technical University of Munich, James
  Franck Street, 85748, Germany}
\address{$^3$ Institute of Materials Research and Engineering, 3 Research
  Link, Singapore, 117602}
\address{$^4$ Department of Physics, National University of Singapore,
2 Science Drive 3, Singapore, 117542}

\ead{christian.kurtsiefer@gmail.com, mengkhoon.tey@gmail.com}

\begin{abstract}
We characterize the interaction between a single atom or similar microscopic
system and a light field via the scattering ratio. For that, we first
derive the electrical field in a strongly focused Gaussian light beam, and then
consider the atomic response. Following the simple scattering model, 
the fraction of scattered optical power for a weak coherent probe field
leads to unphysical scattering ratios above 1 in the strong focusing regime. A
refined model considering interference between exciting and scattered field
into finite-sized detectors or optical fibers is presented, and compared to
experimental extinction measurements for various focusing strengths. 
\end{abstract}

\maketitle

\section{Introduction\label{intro}}
Atom-light interaction at the single quantum level plays an important role in
many  quantum communication and computation protocols. While spontaneous
emission can provide a natural transfer of atomic states into photonic qubits,
strong interaction of light with an atom is needed to transfer a photonic
qubit into internal atomic degrees of freedom as a stationary qubit.
This process is essential to implement quantum light-matter interfaces
\cite{cirac:1997,duan:2001,walls:1990}, unless post-selection techniques are
used \cite{rosenfeld:07}. 

The common approach to achieve this strong interaction pursued for a
long time is to use a high finesse cavity around the atom, in which the
electrical field strength of a single photon is enhanced by multiple
reflections between two highly reflective mirrors, resulting in a high
probability of absorption. Another approach to increase the interaction  
between an atom and a 
single photon is simply to focus the light field of a single photon down to a
diffraction limited area, motivated by the fact that the absorption cross
section of an atom is on the order of the square of the optical
wavelength. Recent theoretical research on this matter predicts that the
absorption probability may reach the maximal value of 100\%
for dedicated focusing geometries \cite{leuchs:2007}. In this paper, we study
the interaction strength between a two-level system and a tightly focused
weak coherent light Gaussian beam, which is simpler to prepare.

Such a system has been theoretically investigated by van Enk and Kimble
\cite{van_Enk:2001} and they concluded that one can expect only a weak
interaction. An experiment on single atom absorption has been carried out
a long time ago in the weak focusing regime \cite{wineland:87}, but recent
experimental results with single molecules \cite{molecule1} and atoms
\cite{our_paper} showed an interaction strength which exceeded these
theoretical predictions by far.
In this paper we extend the original theoretical model such that 
it is applicable in the strong focusing regime and provide experimental data
on the extinction by a single atom for various focusing parameters.
Extrapolating from there, we find that the interaction strength in between
light and an atom can indeed be very strong for realistic focusing geometries.

The paper is organized as follows: In Section~\ref{theory}, we explain how we
quantify the interaction strength between an atom and a weak coherent light
field, and set out the basic problem.  In
Section~\ref{sec:focusfield}, we calculate the field strength at the focus of
an ideal 
lens by considering a Gaussian incident beam for the strong focusing regime. 
Using this `ideal' focusing field developed in this section, we then
obtain an expression for the scattering ratio in Section~\ref{scatt_theory},
and for the extinction of a focused light beam by a two-level system in
various geometries in Section~\ref{sec:extinction}. The theoretical prediction
is compared with our experimental results in Section~\ref{sec:experiment}.

\section{Basic problem\label{theory}}
The system that we investigate is a single two-level atom localized in free
space illuminated by a focused weak monochromatic light field (probe) with 
an incident power $P_\mathrm{in}$. The interaction strength of the probe with
the atom is directly related to the fraction of power scattered by the 
atom. Therefore, it seems reasonable to quantify the interaction strength with
the ratio of the scattered light power $P_\mathrm{sc}$ to the total incident
power $P_\mathrm{in}$, i.e.
\begin{equation}\label{definition}
R_\mathrm{sc}:=\frac{P_\mathrm{sc}}{P_\mathrm{in}}\,.
\end{equation}

\begin{figure}
\begin{center}
  \includegraphics[width=0.65\columnwidth]{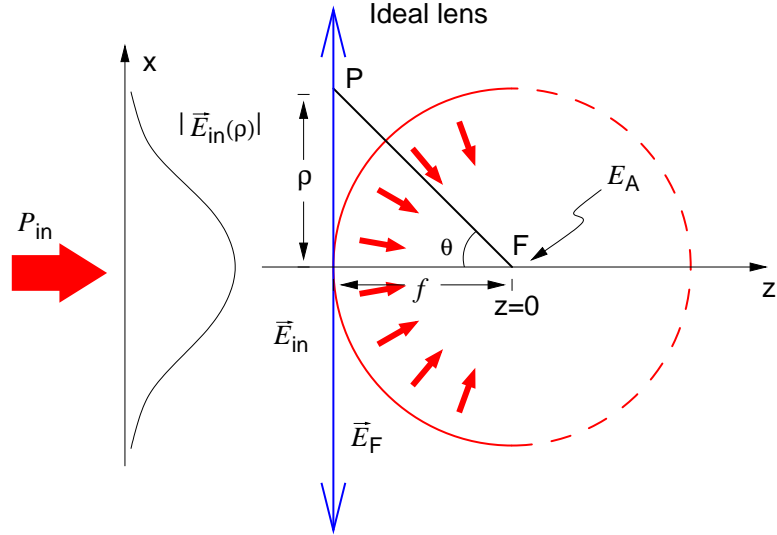}
  \caption{\label{fig:geometry}the electrical field $\vec{E}_\mathrm{in}$ of a
    collimated beam with Gaussian profile is transformed into a focusing field
    $\vec{E}_\mathrm{F}$ with a spherical wave front by an ideal thin lens
    with a focal length $f$, leading to a field amplitude $E_A$ at the
    location of an atom. } 
\end{center}
\end{figure}

To prepare an atom into a clean two-level system, it is convenient to
use optical pumping with circularly polarized light - the optical transition
therefore will also be driven by circularly polarized light. The light field
itself should have a well-defined spatial profile before it is focused onto
the atom with a lens. A circularly polarized, collimated Gaussian beam
propagating along the $z$ axis (see figure~\ref{fig:geometry}) will
therefore be the starting point for our work. Its electrical field strength
before the lens is given by
\begin{equation}
\label{EFS_beforelens}
\vec{E}(\rho, t)=\frac{E_\mathrm{L}}{\sqrt{2}}\left[\cos(\omega
  t)\hat{x}+\sin(\omega t)\hat{y}\right]e^{-\rho^2/w_\mathrm{L}^2}\,,
\end{equation}
where $\rho$ is the radial distance from the lens axis, $w_\mathrm{L}$ the waist of the
beam, $\hat{x}, \hat{y}$ are the unit vectors in transverse directions,
and $E_\mathrm{L}$ is the field amplitude. The beam carries a total power of
\begin{equation}
\label{incidentPower}
P_\mathrm{in}=\frac{1}{4}\epsilon_0\pi cE_\mathrm{L}^2w_\mathrm{L}^2\,,
\end{equation}
where $\epsilon_0$ is the electric permittivity of vacuum, and $c$ the speed
of light in vacuum. Due to the rotational symmetry 
the field on the lens axis is always circularly polarized.
For an atom that is stationary at the focal point of the lens,
the electric field can thus be written as
\begin{equation}
\label{field_at_atom}
\vec{E}(t)=\frac{E_\mathrm{A}}{\sqrt{2}}\left[\cos(\omega
  t)\hat{x}+\sin(\omega t)\hat{y}\right]\,,
\end{equation}
where $E_\mathrm{A}$ denotes the amplitude of the field at the focus. In the
long wavelength limit the atom only interacts with the field at the location
of the atom. 
For a field that is resonant with the atomic transition and with an intensity
much below saturation, the power scattered by a two-level atom is~\cite{cohen-tann:1992}
(see ~\ref{App1} for more details)
\begin{equation}
\label{scattered_power2}
P_\mathrm{sc}=\frac{3\epsilon_0c\lambda^2E_\mathrm{A}^2}{4\pi},
\end{equation}
leading to a scattering ratio of
\begin{equation}
\label{scatt_ratio}
R_\mathrm{sc}=\frac{P_\mathrm{sc}}{P_\mathrm{in}}=\frac{3\lambda^2}{\pi^2w_\mathrm{L}^2}\left(\frac{E_\mathrm{A}}{E_\mathrm{L}}\right)^2\,,
\end{equation}
which is exact under weak and on-resonant
excitation. To evaluate the the scattering ratio $R_\mathrm{sc}$ and therefore
the interaction strength, one needs to know $(E_\mathrm{A}/E_\mathrm{L})^2$. 

For a weakly focused field where the paraxial approximation holds
one finds that
\begin{equation}
\label{paraxial}
\left(\frac{E_\mathrm{A}}{E_\mathrm{L}}\right)^2 \simeq
\left(\frac{w_\mathrm{L}}{w_\mathrm{f}}\right)^2\,,
\end{equation}
where $w_\mathrm{f}$ is the Gaussian beam waist at the focus. This leads to
\begin{equation}
\label{scatt_ratio_parax}
R_\mathrm{sc}\simeq\frac{3\lambda^2}{\pi^2 w_\mathrm{f}^2}=3u^2\,,
\end{equation}
with the focusing strength
\begin{equation}\label{eqdefu}
u:=w_L/f
\end{equation}
used to fix the focal waist, $w_f=\lambda/(\pi u)$. With a Gaussian focal spot
area $A=\pi w_\mathrm{f}^2/2$, the scattering ratio can also be expressed as
$R_\mathrm{sc} \simeq \sigma_\mathrm{max}/A$, where
$\sigma_\mathrm{max}=3\lambda^2/2\pi$ is the absorption cross section of a
two-level system exposed to a resonant plane wave. 
However, for strongly focused light, the paraxial approximation
breaks down, and we need other methods to find $(E_\mathrm{A}/E_\mathrm{L})^2$.

\section{Electrical field in a tight focus\label{sec:focusfield}}
The paraxial approximation breaks down for strongly focused beams both in the
expression of the electric field just behind the lens, and in the propagation
of this field to the focus.
An approach to overcome the propagation problem was reported by van Enk
and Kimble in~\cite{van_Enk:2001}. Their lens model,
however, applies only to the weak focusing regime. In the following, we
present a lens model to overcome this limitation and propagate the optical
field behind the ideal lens into the focal regime and investigate the focal
field using their technique numerically. We obtain a closed
expression for the electrical field in the focus using the Green
theorem for the propagation. 
 
To simplify the expressions in this section, we express the electrical
field in dimensionless units, so the electrical field strength of the
collimated Gaussian beam entering the 
focusing lens is given by
\begin{equation}\label{input_gaussian_field}
\vec{F}_\mathrm{in}=\hat{\epsilon}_+e^{-\rho^2/w_\mathrm{L}^2}\,,
\end{equation}
where $\hat{\epsilon}_+$ is one of the circular polarization vectors
$\hat{\epsilon}_\pm=(\hat{x}\pm i \hat{y})/\sqrt2$.

\subsection{Model of an ideal lens\label{field}}
An ideal converging lens converts a beam with a plane wave front into one with a
spherical wave front which converges towards the focal point $F$. Therefore, it
can be modeled as a phase plate modifying an incoming field $F_\mathrm{in}$
with a radially dependent phase factor  $\varphi(\rho)$ into
\begin{equation}\label{unphysical_field}
\vec{F}_\mathrm{F} = \varphi(\rho)\vec{F}_\mathrm{in}\,.
\end{equation}
In paraxial optics, a convenient analytical treatment of
Gaussian beams can be obtained assuming a parabolic phase factor,
\begin{equation}\label{parabolic_phase_factor}
\varphi_\mathrm{pb}(\rho)=e^{-ik\rho^2/2f}\,;
\end{equation}
this was adopted in \cite{van_Enk:2001}.
However, the conversion of a plane
into a spherical wave front corresponds to
a phase factor of 
\begin{equation}\label{spherical_phase_factor}
\varphi_\mathrm{sp}(\rho)=e^{-ik\sqrt{\rho^2+f^2}}\,,
\end{equation}
which is only approximated by equation (\ref{parabolic_phase_factor}). On top
of this, multiplication of a incoming field with such a phase factor
leads to an electrical field which is not
compatible with Maxwell equations, since the polarization vector for $\rho>0$
is not tangential to the wave front anymore. 

In view of this, we have to change the local polarization with three
requirements in mind  \cite{wolf:1959}:  
(i) A rotationally symmetric lens does not alter the  local azimuthal field
component, but tilts the local radial  polarization component of the
incoming field towards the axis;
(ii) The polarization at point P (see Figure~\ref{fig:geometry}) after
transformation by the lens is orthogonal to the line FP;
(iii) The power flowing into and out of an arbitrarily small area on the thin
ideal lens is the same.
These requirements determine
completely the focusing field right after the lens. With the input field in
equation~(\ref{input_gaussian_field}), one finds (see~\ref{App2} for details)
\begin{eqnarray}\nonumber
\fl \vec{F}_\mathrm{F}(\rho,\phi,z=-f)=&\frac{1}{\sqrt{\cos\theta}}\textrm{
}\left(\frac{1+\cos\theta}{2}\textrm{}\hat{\epsilon}_+ + \frac{\sin\theta
    e^{i\phi}}{\sqrt{2}}\textrm{}\hat{z}+\frac{\cos\theta-1}{2}e^{2i\phi}\textrm{}\hat{\epsilon}_-\right)\\&
\times \exp\left(-\rho^2/w_\mathrm{L}^2\right)\textrm{
}\exp\left[-ik\sqrt{\rho^2+f^2}\right]\,, 
\label{physical_focused_field}
\end{eqnarray}
with $\theta=\mathrm{arctan} (\rho/f)$. In particular, the factor $1/\sqrt{\cos\theta}$
is needed in order to meet requirement (iii).

\subsection{Numerical propagation of the field to the focus}
The optical field with a converging wave front directly behind the lens needs
to be propagated into the focal region to arrive at a field strength of the
light interacting with the microscopic system. Various methods can be applied
for this purpose. The one implemented in \cite{van_Enk:2001} projects the
focusing field $\vec{F}_\mathrm{F}$ on an orthogonal set of modes $\vec{F}_\mu$,
$\mu=(k_t,s,m)$ with cylindrical symmetry (see \ref{sec:cylmodes} for
details). This decomposition reads
\begin{equation}
\vec{F}_\mathrm{F} =\sum\limits_\mu\kappa_\mu\vec{F}_\mu
\end{equation}
where the expansion coefficients $\kappa_\mu$ are given by
\begin{eqnarray}\nonumber
\fl \kappa_\mu=\delta_{m1}\pi k_t \int_0^\infty d\rho\textrm{ }\rho\frac{1}{\sqrt{\cos\theta}}\textrm{
}\textrm{{\Large \{}}
\frac{sk+k_z}{k}\left(\frac{1+\cos\theta}{2}\right)J_0(k_t \rho)+i
\frac{\sqrt{2} k_t}{k}\left(\frac{\sin\theta}{\sqrt{2}}\right) J_1(k_t
\rho)\\&\fl\fl\fl\fl\fl+\frac{sk-k_z}{k}\left(\frac{\cos\theta-1}{2}\right)
J_2(k_t\rho)\textrm{{\Large
    \}}}\exp\left[-ik\sqrt{\rho^2+f^2}\,-\frac{\rho^2}{w_\mathrm{L}^2}\right]\,, 
\end{eqnarray}
with $\theta=\tan^{-1}(\rho/f)$. The Kronecker symbol $\delta_{m1}$ 
reflects conservation of angular momentum under the lens transformation
\cite{vanenk:1992,beijersbergen:1993}. The projection integral has no analytic
solution, so the coefficients have to be evaluated numerically.

The (dimensionless) field components in the three polarization components
$\hat\epsilon_\pm, \hat{z}$ at any point behind the lens are
superpositions of contributions from different modes,
\begin{eqnarray}\label{physical_F+}
F_+(\rho, \phi, z)&=&\sum_{s=\pm1}\int_{0}^{k}dk_t\textrm{ }\frac{1}{4\pi}\frac{sk+k_z}{k}J_0(k_t\rho)e^{i k_z z} \kappa_{\mu},\\\label{physical_Fz}
F_z(\rho, \phi, z)&=&\sum_{s=\pm1}\int_{0}^{k}dk_t\textrm{ }(-i)\frac{\sqrt{2}}{4\pi}\frac{k_t}{k} J_1(k_t\rho)e^{i k_z z}e^{i\phi} \kappa_{\mu},\\\label{physical_F-}
F_-(\rho, \phi, z)&=&\sum_{s=\pm1}\int_{0}^{k}dk_t\textrm{
}\frac{1}{4\pi}\frac{sk-k_z}{k}J_2(k_t\rho)e^{i k_z z}e^{2i\phi}
\kappa_{\mu}.\end{eqnarray} 
We now evaluate the field components for different regions with this method:
first directly behind the lens, then on the optical axes to near the focus,
and finally in the focusing plane near the focus.

\subsubsection{Focusing field reconstruction}
As a consistency check, we first evaluate the field components right after the
lens (i.e., for $z=-f$)  for a reasonably strong focusing field with $u=1.56$
corresponding to $w_\mathrm{L}=7$\,mm for $f=4.5$\,mm. The relative difference between 
the reconstructed and original field is less than $10^{-3}$, a bound limited by
our numerical accuracy. A linear combination of the field modes $\mu$ in the
form of equations (\ref{physical_F+}) to (\ref{physical_F-}) is compatible
with the Maxwell equations, so since there is no significant difference
between the original and reconstructed field, the choice in equation
(\ref{physical_focused_field}) for the focusing field is
compatible with Maxwell equations even for strong focusing parameters.

\subsubsection{Field along the optical axis}
\begin{figure}
  \begin{center}
    \includegraphics[width=0.85\columnwidth]{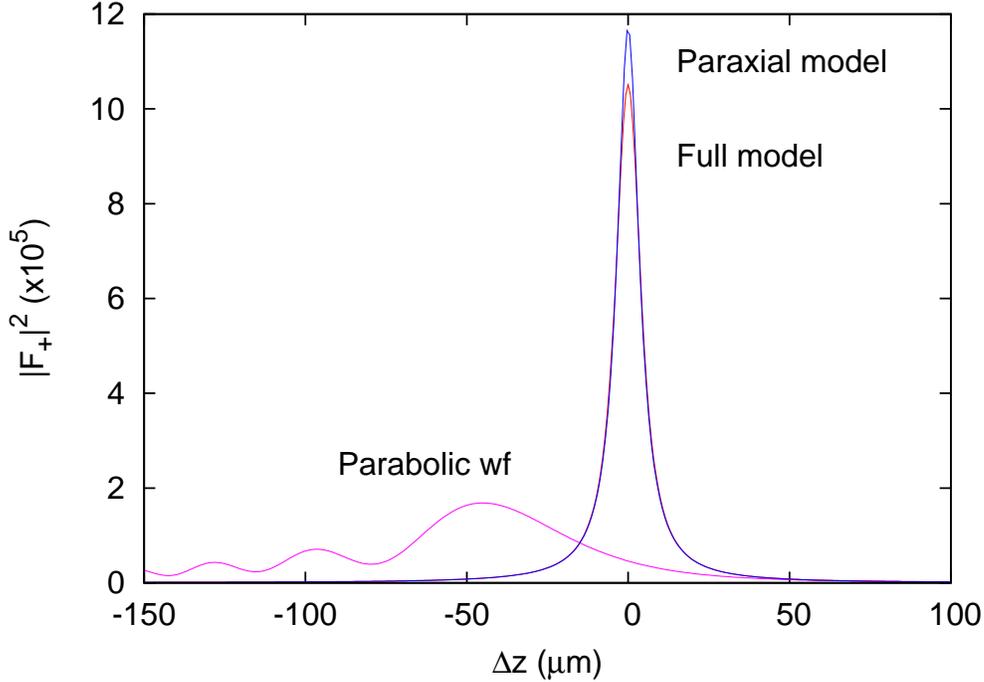}
    \caption{Dimensionless intensity $|F_+|^2$ along the optical axis for the
      full focusing field model according to equation
      (\ref{physical_focused_field}), the paraxial Gaussian beam model, 
      and for comparison with a parabolic phase factor for the lens only.
      The latter leads to spherical aberration, manifesting in a spread
      of the focal area and a shift towards the lens. }
    \label{fig:field_comparison}
  \end{center}
  \end{figure}

The solid line in figure~\ref{fig:field_comparison} shows the 
dimensionless intensity $|F_+|^2$ for $f=4.5$\,mm, $\lambda=780$\,nm and
$w_\mathrm{L}=1.1$\,mm ($u=0.244$), with a clearly peaked distribution
centered at the focus $\Delta z=0$ and a depth of field, defined as the full width at half
maximum (FWHM), of about 9.5\,$\mu$m. This result is still very close to the
much simpler paraxial approximation of a Gaussian beam, with a depth of field of
$2\lambda/(\pi u^2)=8.31\,\mu$m.

For comparison, we show the result for a focusing field using a
parabolic phase factor $\varphi_\mathrm{pb}$ only, following
\cite{van_Enk:2001}.
The spherical aberration there displaces and spreads the focus, and
significantly reduces the maximal intensity in the focal point $F$. This
problem becomes even more serious for a larger input waist $w_\mathrm{L}$.

\subsubsection{Field in the focal plane\label{field_at_focus}}
\begin{figure}
  \begin{center}
    \includegraphics[width=\columnwidth]{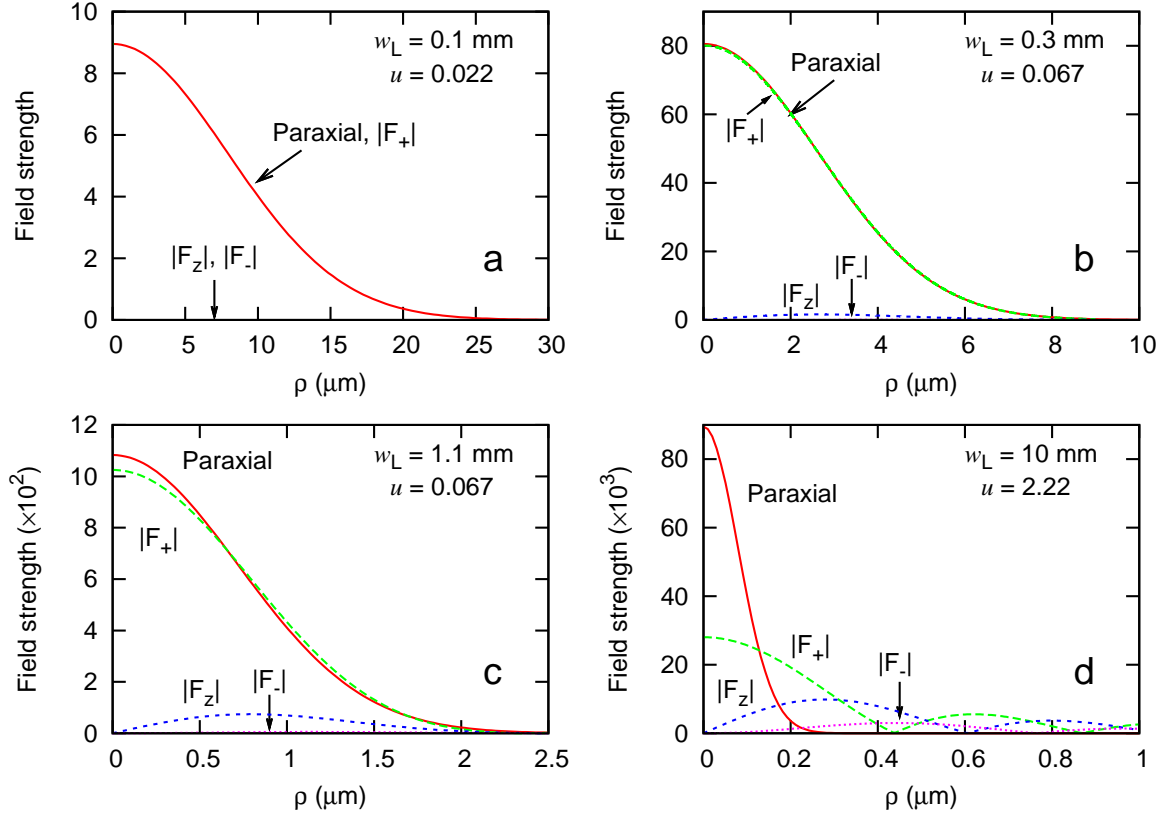}\\
    \caption{Field amplitudes at the focus for different focusing
      strengths. All plots are for a focal length of 4.5\,mm and wavelength of
      780\,nm.}
    \label{fig:field_at_focus_4graphs}
  \end{center}
\end{figure}

We now examine the field near the focus in more details. Figure  
\ref{fig:field_at_focus_4graphs} shows the field on the focal plane for 
different focusing strengths. For this, we choose different 
input waists $w_\mathrm{L}$, but keep $f$=4.5\,mm and $\lambda$=780\,nm fixed.
For comparison, we also show the result for a focusing field according to the
paraxial approximation, 
\begin{equation}\label{focal_gaussian_field}
\vec{F}_\mathrm{parax}=\frac{w_\mathrm{L}}{w_\mathrm{f}}\hat{\epsilon}_+e^{-\rho^2/w_\mathrm{f}^2},
\end{equation}
with a paraxial focal waist $w_\mathrm{f}=f\lambda/\pi w_\mathrm{L}$.  
For weak focusing ($u=0.022, w_\mathrm{L}=0.1\,$mm) in
figure~\ref{fig:field_at_focus_4graphs}a, $|F_+|$ 
overlaps completely with the paraxial prediction with negligible $|F_z|$ and
$|F_-|$. For an initial waist $w_\mathrm{L}=0.3$\,mm corresponding to
$u=0.067$ and $w_\mathrm{f}\simeq \textrm{3.7} \mu\textrm{m}$ (about
5$\lambda$),  
discrepancies between the paraxial approximation and the extended model start to
appear (Figure~\ref{fig:field_at_focus_4graphs}b). With increasing
$w_\mathrm{L}$, the $\hat{z}$- and $\hat{\epsilon}_-$ polarized  field
components become stronger for $\rho>0$, but an atom localized on the 
optical axis still only experiences a
$\hat{\epsilon}_+$-polarized field.  Figure~\ref{fig:field_at_focus_4graphs}d
shows the focused field that maximizes $|F_+|$ for the parameters in our
model. It is obtained with an incident waist  $w_\mathrm{L}=10$\,mm
($u=2.22$). An 
increase of the incident waist beyond that does not reduce the focal spot size
any further due to the diffraction limit. Instead, more energy is transferred
to the other polarization components, thus decreasing the magnitude of the
$F_+$. 

\subsection{Analytical expression for the field in the focal point}
An alternative method for propagating the focusing field directly behind the
lens is offered by the Green theorem. For given electrical
and magnetic fields $\vec{E}(\vec{r}\,')$ and $\vec{B}(\vec{r}\,')$ on an
arbitrary closed surface $S'$ that encloses a point $\vec{r}$, the
electrical field at this point is determined by \cite{jackson:CE}
\begin{eqnarray}\nonumber
\vec{E}(\vec{r})=\oint_{S'}\mathrm{ }dA'\textrm{ }&&\textrm{{\Large \{}}ikc\left[\vec{n}\,'\times\vec{B}(\vec{r}\,')\right]G(\vec{r},\vec{r}\,')+\left[\vec{n}\,'\times\vec{E}(\vec{r}\,')\right]
\times\nabla'G(\vec{r},\vec{r}\,')\\ && +\left[\vec{n}\,'\cdot\vec{E}(\vec{r}\,')\right]\nabla'G(\vec{r},\vec{r}\,')\textrm{{\Large \}}},\label{green_theorem}
\end{eqnarray}
where $\vec{n}\,'$ is the unit vector normal to a differential surface
element $dA'$ and points into the volume enclosed by $S'$, and $G(\vec{r},\vec{r}\,')$ is the
Green function given by 
\begin{equation}
G(\vec{r},\vec{r}\,')=\frac{e^{ik|\vec{r}-\vec{r}\,'|}}{4\pi |\vec{r}-\vec{r}\,'|}\,.
\end{equation}

If point $\vec{r}$ is the focus of an aplanatic focusing field, then the local field propagation wave vector $\vec{k}\,'$ at any point $\vec{r}\,'$ always points towards (away from) point $\vec{r}$ for the incoming (outgoing) field in the far field limit, i.e. when $|\vec{r}-\vec{r}\,'|\gg\lambda$. In this limit, one has
\begin{eqnarray}
B(\vec{r}\,')\rightarrow \frac{\vec{k}\,'}{c|\vec{k}\,'|}\times E(\vec{r}\,'),\label{BE_relation} \\
\nabla'G\rightarrow -i\vec{k}\,'G\,\textrm{ before the focus}\,,
\nabla'G\rightarrow i\vec{k}\,'G\,\textrm{ after the focus}\,.\label{gradientapprox}
\end{eqnarray}
In the far field limit, \Eref{green_theorem} reduces to
\begin{eqnarray}\nonumber
\vec{E}(\vec{r}_{\mathrm{\,\,focus}})&=&-2i\int_{S_\mathrm{bf}} dA'\textrm{ }\left[\vec{n}\,'\cdot\vec{k}\,'\right]\vec{E}(\vec{r}\,')G(\vec{r},\vec{r}\,')\\&&+2i\int_{S_\mathrm{af}} dA'\textrm{ }\left[\vec{n}\,'\cdot\vec{E}(\vec{r}\,')\right]\vec{k}\,'G(\vec{r},\vec{r}\,')\,.\label{simplified_Green}
\end{eqnarray}
Here the surface $S'$ is divided into two parts, where $S_\mathrm{bf}$ is the
one before the focal plane, and $S_\mathrm{af}$ is the
surface after the focal plane. The second term in \Eref{simplified_Green} vanishes if we choose $S_\mathrm{af}$ to be an infinitely large hemisphere
centered at the focus, since in this case $\vec{n}\,'$ is perpendicular to
$\vec{E}(\vec{r}\,')$ at all points on $S_\mathrm{af}$ for an aplanatic
field. If we choose $S_\mathrm{bf}$ as an infinitely large plane that
coincides with the ideal lens and adopt the dimensionless incident field in
\Eref{physical_focused_field}, we get 
\begin{equation}\label{focus_inte}
\vec{F}(0,0,z=0)=\frac{-ik\sqrt{f}}{2}\int_0^\infty d\rho\,\frac{\rho(f+\sqrt{f^2+\rho^2})}{(f^2+\rho^2)^{5/4}}\exp(-\frac{\rho^2}{w_L^2})\,\hat{\epsilon}_+\,
\end{equation}
which has an analytical solution 
\begin{equation}\label{focus_solution}
\vec{F}(0,0,z=0)=-\frac{1}{4}{ikw_L\over u} e^{1/u^2}\left[\sqrt{1\over
    u}\Gamma(-{1\over4},{1\over u^2})+\sqrt{u}\Gamma({1\over4},{1\over u^2})\right]\,\hat{\epsilon}_+\,,
\end{equation}
with the incomplete gamma function $\Gamma(a,b)=\int_b^\infty\,t^{a-1}e^{-t}\,dt\,$, and $u=w_L/f$ as
in \Eref{eqdefu}. The results obtained with the mode decomposition method agree with this
expression within computational errors of about 0.1\%. 
The $-i$ reflects a Gouy phase of $-\pi/2$ \cite{born:PO}.

We now restore the field dimensions by multiplication with
the amplitude in the center of the collimated Gaussian beam, which can
be expressed by the optical power according to equation (\ref{incidentPower}),
\begin{equation}
E_L={1\over w_L}\sqrt{4P_\mathrm{in}\over\epsilon_0\pi c}\,,
\end{equation}
resulting in an electrical field amplitude in the focus of 
\begin{equation}
\left|E_A\right|=\sqrt{\pi P_\mathrm{in}\over\epsilon_0 c\lambda^2}\,\cdot\,{1\over u} e^{1/u^2}\left[\sqrt{1\over
    u}\Gamma(-{1\over4},{1\over u^2})+\sqrt{u}\Gamma({1\over4},{1\over u^2})\right]\label{eq:absolute_focus_field}
\end{equation}
with purely circular polarization. The focal field thus only depends on
the input power, the optical wavelength and the focusing
strength $u$. 

\section{Study of the scattering ratio\label{scatt_theory}}
\begin{figure}
\begin{center}
  \includegraphics[width=0.85\columnwidth]{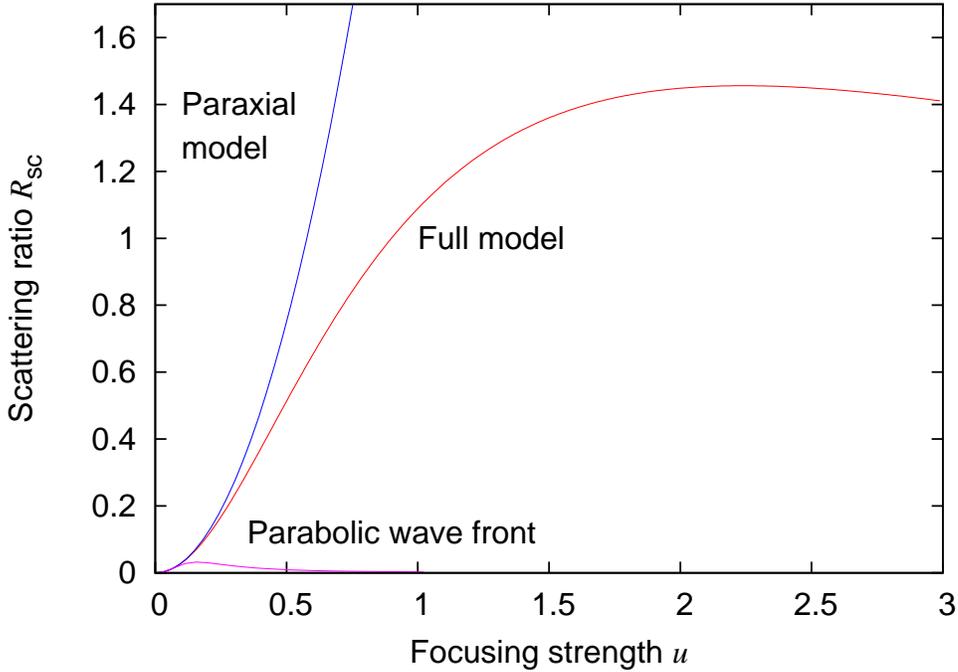}
  \caption{Scattering ratio $R_\mathrm{sc}$ as a function of the focusing
    parameter $u=w_\mathrm{L}/f$ with the full focusing field, in paraxial
    approximation and with a parabolic wavefront.} 
  \label{fig:scat_prob_4models}
\end{center}
\end{figure}

The electrical field amplitude $E_A$ at the focus for a given excitation power
now allows to determine the fraction $R_\mathrm{sc}$ of the optical power scattered
away by a two-level atom or similar microscopic object according to
equation (\ref{scattered_power2}). With \Eref{eq:absolute_focus_field},
we arrive at
\begin{equation}
R_\mathrm{sc}={3\over4u^3}\, e^{2/u^2}\left[\Gamma(-{1\over4},{1\over u^2})+u\Gamma({1\over4},{1\over u^2})\right]^2\,.\label{eq:absolute_scatteringrate}
\end{equation}

In Figure~\ref{fig:scat_prob_4models} we show this quantity as a function of the focusing strength $u$. A striking feature of this plot is that, for $u$ large enough, $R_\mathrm{sc}$ exceeds the value of 1, as if more light would be scattered than was incident. However, $R_\mathrm{sc}$ cannot be interpreted as a scattering ratio anymore if the solid angle subtended by the excitation field is not negligible: in the strong focusing regime, interference between the exciting field and the scattered field must be taken into account. In fact, the physical bound (Bassett limit) is $R_\mathrm{sc}\leq 2$~\cite{bassett:1986}. For our focusing model of the Gaussian beam, we predict a maximal value of $R_\mathrm{sc}=1.456$ for a focusing strength $u=2.239$.

For reference, we also show $R_\mathrm{sc}$ for focal fields derived under the paraxial approximation and for a parabolic wave front model. All models agree in the weak focusing regime, and $R_\mathrm{sc}$ can reach values on the order of 1, which indicates that by strong focusing, one can accomplish a strong interaction between a light field and a single atom.

\section{Extinction as a measurement of scattering\label{sec:extinction}}

The parameter $R_\mathrm{sc}$ we have just discussed can be used as a figure-of-merit for scattering experiments. In this section, we go beyond this one-parameter description to provide a detailed model for the experiment we performed \cite{our_paper}. In this experiment, we measured the extinction of the transmitted light beam due to scattering.

\begin{figure}
  \begin{center}
    \includegraphics[width=0.65\columnwidth]{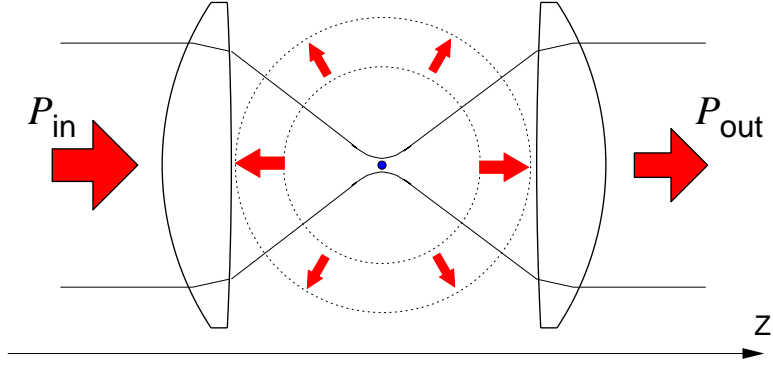}
    \caption{A transmission measurement setup with an atom at the focus of a lens. The transmitted power is a result of interference between the scattered light and the probe for coherent scattering.\label{img:idea}}
\end{center}
\end{figure}

Figure~\ref{img:idea} illustrates a simple transmission setup with an atom
located at the focus of two confocal lenses, where
the second lens collects all the excitation power if no atom is
present at the focus. The actual measured transmission $T$ depends on the area
covered by the power detector after the second lens, and can be obtained by
considering the interference between the incident probe field and the field
coherently scattered by the atom. For a given focusing and collection geometry,
the extinction  
\begin{equation}
\epsilon=1-T={P_\mathrm{in}-P_\mathrm{out}\over P_\mathrm{in}}
\end{equation}
is maximal for a weak incident probe field in resonance with the transition in
the two-level system. The total field at any place is a superposition of the
focusing field exciting the atom, and the scattered field:
\begin{equation}\label{totalfield}
\vec{E}_\mathrm{t}(\vec{r})= \vec{E}_\mathrm{F}(\vec{r})+\vec{E}_\mathrm{sc}(\vec{r})\,.
\end{equation}
The spatial dependency of the scattered field $\vec{E}_\mathrm{sc}$ is that of
a rotating electrical dipole, with an amplitude proportional to the exciting
electrical field amplitude $E_A$, and the total power contained in this dipole
radiation must match \Eref{scattered_power2}. Far away from the dipole
($r\gg\lambda$), this scattered field takes the form
\begin{equation}\label{dipole_field}
\vec{E}_\mathrm{sc}(\vec{r})=\frac{3E_Ae^{i(kr+\pi/2)}}{2kr} 
\left[\hat{\epsilon}_+-(\hat{\epsilon}_+\cdot\hat{r})\,\hat{r}\right]\,,
\end{equation}
where $\hat{r}$ is the radial unit vector pointing away from the 
scatterer~\cite{cohen-tann:1992}. The $\pi/2$ phase reflects the fact that the
dipole moment of the atom lags the field $E_A$  by $\pi/2$ at resonance.
The focusing field $\vec{E}_\mathrm{F}$ close to the lenses at $z=\pm f$ takes
the form
\begin{eqnarray}
\fl \vec{E}_\mathrm{F}(\rho,\phi,z=\pm f)=&
\frac{E_L}{\sqrt{|\cos\theta}|}\textrm{}
\left(\frac{1\mp\cos\theta}{2}\textrm{}\hat{\epsilon}_+ 
\mp \frac{\sin\theta e^{i\phi}}{\sqrt{2}}\textrm{}\hat{z}+
\frac{\mp\cos\theta-1}{2}e^{2i\phi}\textrm{}\hat{\epsilon}_-\right)\nonumber\\&
\times \exp\left(-\rho^2/w_\mathrm{L}^2\right)\textrm{ }
\exp\left[\pm i(k\sqrt{\rho^2+f^2}-\pi/2)\right]\,, 
\label{physical_focused_field2}
\end{eqnarray}
where $\theta\in[0,\pi]$ is the polar angle between the $-z$ direction and a
point $(\rho, \phi, z)$ as in figure~\ref{fig:geometry}. The phase is
adjusted such that the electric field amplitude $E_A$ at the focus is real. 

The excitation
field and the forward scattered field interfere destructively, as was shown
first for the case of an incident plane
wave~\cite{bohren:1983,paul:1983,davis:2001}, and more recently for arbitrary
incident fields \cite{zumofen:2008} with the help of vectorial multipole
expansions \cite{bohren:ASLSP, sheppard:1997,mojarad:2008}.

\subsection{Energy flux through transverse planes}
The optical power $P_\mathrm{out}$ arriving at a detector behind the atom
or a collection lens can be evaluated from the superposition of fields via the
time averaged energy flux through a plane $S$ with a fixed $z$,
\begin{equation}
P_S = \frac{\epsilon_0 c^2}{2}\int_S
\Re\left\{\vec{E}_\mathrm{t}\times\vec{B}^*_\mathrm{t}\right\}\cdot d\vec{A}
\end{equation}
where $d\vec{A}$ is a differential area element of the surface $S$ and
$\Re(x)$ denotes the real part of $x$.

Far away from the focus, the electromagnetic field can be locally approximated
by a plane wave such that $\vec{B}=\hat{k}\times\vec{E}/c$, where
$\hat{k}=\hat{k}_\mathrm{sc},\,\hat{k}_\mathrm{F}$ is a dimensionless unit
vector parallel to the local field propagation direction. Both
$\vec{E}_\mathrm{F}$ and $\vec{E}_\mathrm{sc}$ have spherical wave
fronts, i.e., the local propagation directions are parallel. Before the focus
we have $\hat{k}_\mathrm{sc}=-\hat{k}_\mathrm{F}$, while after the focus we
have $\hat{k}_\mathrm{sc}=\hat{k}_\mathrm{F}$. With these field properties,
and with the local 
transversality, $\hat{k}_\mathrm{F}\cdot\vec{E}_\mathrm{F}=0$ and
$\hat{k}_\mathrm{F}\cdot\vec{E}_\mathrm{sc}=0$, the power through the two
planes can be expressed with electrical fields only,
\begin{equation}
\fl \textrm{     }P_\mathrm{z=\pm f}=\frac{\epsilon_0 c}{2}
\int_\mathrm{z=\pm f}\Re\left\{\vec{E}_\mathrm{F}\cdot\vec{E}_\mathrm{F}^*\pm\vec{E}_\mathrm{sc}\cdot\vec{E}_\mathrm{sc}^*+\vec{E}_\mathrm{sc}\cdot\vec{E}_\mathrm{F}^*\pm\vec{E}_\mathrm{F}\cdot\vec{E}_\mathrm{sc}^*\right\}\hat{k}_\mathrm{F}\cdot\hat{z}dA\label{flux_at_pmf}\,.
\end{equation}
The two first terms represent (i) the power of the excitation field, (ii) the power of the scattered field,
while the third and fourth term represent (iii) the interference term.

The contribution (i) to $P_\mathrm{z=\pm f}$ is simply the input power,
\begin{equation}\label{Uf_in}
P_\mathrm{z=\pm f,\,in}:=\frac{\epsilon_0 c}{2}
\int_\mathrm{z=\pm
  f}\Re\left\{\vec{E}_\mathrm{F}\cdot\vec{E}_\mathrm{F}^*\right\}\hat{k}_\mathrm{F}\cdot\hat{z}dA
= \frac{1}{4}\epsilon_0\pi cE_\mathrm{L}^2w_\mathrm{L}^2=P_\mathrm{in}\,,
\end{equation}
and the contribution (ii) in these planes,
\begin{equation}
\left|P_\mathrm{z=\pm f,\,sc}\right| :=\frac{\epsilon_0 c}{2}
\int_\mathrm{z=\pm f}\Re\left\{\vec{E}_\mathrm{sc}\cdot\vec{E}_\mathrm{sc}^*\right\}\hat{k}_\mathrm{F}\cdot\hat{z}dA
= \frac{3\epsilon_0c\lambda^2E_\mathrm{A}^2}{8\pi}=\frac{P_\mathrm{sc}}{2}\,,\label{Uf_sc}
\end{equation}
where $P_\mathrm{sc}$ is the scattered power as defined previously in
\Eref{scattered_power2}.

The interference contribution (iii) vanishes for $z=-f$ because
$\left(\vec{E}_\mathrm{sc}\cdot\vec{E}_\mathrm{F}^*-\vec{E}_\mathrm{F}\cdot\vec{E}_\mathrm{sc}^*\right)$
is purely imaginary, whereas for $z=+f$ we get
\begin{eqnarray}\label{Uf_int}
P_\mathrm{z=+f,\,int}&:=\frac{\epsilon_0 c}{2}
\int_\mathrm{z=\pm f}\Re\left\{\vec{E}_\mathrm{sc}\cdot\vec{E}_\mathrm{F}^*+\vec{E}_\mathrm{F}\cdot\vec{E}_\mathrm{sc}^*\right\}\hat{k}_\mathrm{F}\cdot\hat{z}dA\\
&=-\frac{3\pi\epsilon_0 c E_A E_L \sqrt{f}}{2k}
\int_0^\infty\frac{\rho(f+\sqrt{f^2+\rho^2})}{(f^2+\rho^2)^{5/4}}\exp(-\frac{\rho^2}{w_L^2}) d\rho\,.
\end{eqnarray}
The negative sign, which comes from both the Gouy phase in the incident field
(\Eref{physical_focused_field2}) and the phase difference between the dipole
and local field (\Eref{dipole_field}), reveals that the scattered light and
the incident light interfere destructively after the focus
\cite{zumofen:2008}. This integral can be solved in the same way as 
\Eref{focus_inte}, leading to  
\begin{equation}\label{Uf_int_result}
P_\mathrm{z=+f,\,int} =
-\frac{3\epsilon_0c\lambda^2E_\mathrm{A}^2}{4\pi}=-P_\mathrm{sc}\,.
\end{equation}

Thus, the power flowing through both planes $z=\pm f$ is the same,
\begin{equation}
P_\mathrm{z=\pm f}=P_\mathrm{in}-\frac{P_\mathrm{sc}}{2}\,.
\end{equation}
This indicates what we mentioned above: a value $R_\mathrm{sc}>1$ does not violate energy
conservation, the physical bound being rather $R_\mathrm{sc}\leq 2$.

For later on, we should define a measurable extinction as the difference of the
transmitted power with and without the atom, divided by the power transmitted 
without the atom at a location behind the atom, e.g. at $z=+f$:
\begin{eqnarray}
\epsilon&=\frac{P_\mathrm{z=+f,\,in}-P_\mathrm{z=+f}}{P_\mathrm{z=+f,\,in}}
\nonumber\\
&=\frac{-P_\mathrm{z=+f,\,int}-P_\mathrm{z=+f,\,sc}}{P_\mathrm{z=+f,\,in}}\,.\label{extinction}
\end{eqnarray}
If the collection lens at $z=+f$ is infinitely large, it takes a value of 
\begin{equation}
\epsilon=\frac{P_\mathrm{sc}/2}{P_\mathrm{in}}=\frac{R_\mathrm{sc}}{2}\,.
\end{equation}

\subsection{Extinction observed with a detector/lenses with  finite diameter}
Realistic lenses will have a finite size, thus only partly transmit the
excitation 
and scattered light. We now estimate how this obstruction affects the relation
between an observed extinction and the inferred scattering ratio
$R_\mathrm{sc}$.  

The first effect of a finite lens aperture radius $\rho_0$ of the first lens
is a reduction of the field at the focus. The Green theorem method for
evaluating the focal field via equation (\ref{focus_inte}) still has a closed
solution for a finite radius,

\begin{eqnarray}
  \fl\frac{E_A^{\rho_\mathrm{0}}}{E_L}=
  kf\sqrt{u}e^{1/ u^2}
  \left\{{1\over4}\left[\Gamma\left({1\over4},{1\over u^2}\right)
      -\Gamma\left({1\over4},{1+v^2\over u^2}\right)\right]
    +{1\over u}\left[\Gamma\left(\frac{3}{4},{1+v^2\over u^2}\right)
      -\Gamma\left(\frac{3}{4},{1\over u^2}\right)\right]\right.\nonumber\\
\left. +\sqrt{1\over u}e^{-{1/ u^2}}
  \left[1-\frac{e^{-v^2/ u^2}}{\left(1+v^2\right)^{1/4}}\right]\right\}\label{finite_lens_focus_solution}  
\end{eqnarray}
with $u=w_L/f$ as in \Eref{eqdefu}, and similarly $v:=\rho_\mathrm{0}/f$ as half the
f-number of the lens which is related to the numerical aperture
NA of the lens via 
$\mathrm{NA}^2=v^2/(1+v^2)$.
This obstruction reduces the scattered power $P_\mathrm{sc}$. For realistic
lens sizes, however, this is a very small effect. For instance, for
$\rho_\mathrm{0}=2w_\mathrm{L}$ we find
$P_\mathrm{sc}^\mathrm{\rho_\mathrm{0}}\simeq
0.97P_\mathrm{sc}^\mathrm{\,\infty}$. 

Without the atom, the transmitted power after the collection lens is given by
\begin{equation}
P_\mathrm{f=+z,in}^\mathrm{\rho_\mathrm{0}}=P_\mathrm{in}\left[1-\exp(-\frac{2\rho^2_\mathrm{0}}{w_\mathrm{L}^2})\right]\,.
\end{equation}
The transmitted power in absence of an atom is thus very close to
$P_\mathrm{in}$ for $\rho_\mathrm{0}>2w_\mathrm{L}$. The contribution (iii) of
the interference terms to the forward power is now given by  
\begin{equation}
P_\mathrm{z=+f,\,int}^{\rho_\mathrm{0}}=-P_\mathrm{sc}^\mathrm{\rho_\mathrm{0}}\,,
\end{equation}
and the contribution (ii) of the scattered light by 
\begin{eqnarray}\label{collected_scattered_light}
P_\mathrm{z=+f\,,sc}^\mathrm{\rho_\mathrm{0}} &=& 
{1-\alpha\over2}\cdot P_\mathrm{sc}^{\rho_\mathrm{0}}\\
\mathrm{with}\quad
\alpha&=&\frac{1+3v^2/4}{\left(1+v^2\right)^{3/2}} =
(1-{\mathrm{NA}^2\over4})\sqrt{1-\mathrm{NA}^2}\,. \label{pickup_factor}
\end{eqnarray}
The extinction measurable with a finite aperture lens thus is
\begin{equation}
\epsilon={1+\alpha\over2}
\cdot \frac{P_\mathrm{sc}^{\rho_\mathrm{0}}}{P_\mathrm{in}}
\cdot {1\over1-e^{-2\rho^2_\mathrm{0}/{w_\mathrm{L}^2}}}
\,.\label{exact_extinction}
\end{equation}
If the lenses fully accommodate the Gaussian incident beam, say $\rho_0>2w_L$,
then this can very well be approximated by
\begin{equation}
\epsilon={1+\alpha\over2}\cdot R_\mathrm{sc}^{\rho_\mathrm{0}}\,,
\end{equation}
where
$R_\mathrm{sc}^{\rho_\mathrm{0}}=P_\mathrm{sc}^{\rho_\mathrm{0}}/P_\mathrm{in}$.

We can also quantitatively evaluate the reflectivity of a single atom in this
strong focusing regime, which was recently found to be possibly very large
\cite{zumofen:2008}. The scattered power recollected by the input lens is also
given by  \Eref{collected_scattered_light}, thus the 'single atom
reflectivity' with a Gaussian beam profile is 
\begin{equation}\label{reflectivity}
R= {1-\alpha\over2}\cdot R_\mathrm{sc}^{\rho_\mathrm{0}}\,.
\end{equation}

We conclude that the measured extinction presents a lower
bound to the scattering ratio if the collection lens fully collects the probe
after the focus. For a small numerical aperture of the collection
lens, $\epsilon\simeq R_\mathrm{sc}$ as expected, whereas for a large
collection aperture (corresponding to a small loss factor $\alpha$), a reduced
extinction $\epsilon\rightarrow R_\mathrm{sc}/2$ should be observed.

\subsection{Extinction observed with a detector behind a single mode fiber}  
The symmetrical arrangement of the focusing and collection lens, and the
typical preparation of a Gaussian excitation beam by an optical fiber suggests
that the light could also be collected by a single mode optical fiber.
The confinement of the light field into waveguides
with well-defined mode functions makes the atom-focusing arrangement an
independent building block for 'processing' electromagnetic fields.

The amplitude $a_c$ of the light field collected into the optical fiber is
again given 
by the sum of the excitation and scattered field, picked up by the
optical fiber. This amplitude $a_c$ can be obtained by projecting the field
$\vec{E}_\mathrm{t}$ onto the field mode $\vec{g}$ of the optical fiber. This
projection can be carried out with the scalar product
\begin{equation}
a_c=\left \langle \vec{g}, \vec{E}_\mathrm{t}\right \rangle:={\epsilon_0 c\over2}\int_{\vec{x}\in
  S}\left\{
  \vec{g}^*(\vec{x})\cdot\vec{E}_\mathrm{t}(\vec{x})\right\}
(\hat{k}_{\vec{g}}\cdot\hat{n})\, dA\,, \label{eq:scalprod}
\end{equation} 
where the integration plane $S$ is chosen such that both $\vec{g}$ and
$\vec{E}$ are far away from a focus, $\hat{k}_{\vec{g}}$ is the local propagation
direction of the mode function $\vec{g}$, and $\hat{n}$ the normal vector on
the plane $S$.  This integration can be carried out at any convenient location  
as long as it captures the mode function.
The scalar product in \Eref{eq:scalprod} is written such that it
resembles the form of the power integral in planes $z=\pm f$ in
\Eref{flux_at_pmf}, so we can conveniently use the integrations carried
out earlier. Thus, the integration plane $S$ is chosen at $z=+f$,
directly before the collection lens.

In the experiment, the excitation mode is matched to the collecting
single mode fiber. Correspondingly, we define the target mode function $\vec{g}(x,y,z)$ to be the same
as that of the excitation mode of $\vec{E}_\mathrm{F}$ in equation
(\ref{physical_focused_field2}). With the normalization condition $\left \langle
  \vec{g}, \vec{g}\right\rangle=1$ we simply can set
\begin{equation}
\vec{g}=\vec{E}_\mathrm{F}/\sqrt{P_\mathrm{in}}\,.
\end{equation}
With this normalization, the square of the projection coefficient $a_c$ has the
dimension of a power. Thus, the optical power of the field coupled into the
fiber with a scattering atom present is given by 
\begin{eqnarray}
  P_\mathrm{out}&=&\left| \left \langle \vec{g}, \vec{E}_\mathrm{t} \right
    \rangle \right|^2= \left| \left \langle \vec{g}, \vec{E}_\mathrm{F}+
    \vec{E}_\mathrm{sc} \right \rangle \right|^2 \nonumber \\
&=&\left|\left \langle \vec{g},  \vec{E}_\mathrm{F} \right \rangle 
+\left \langle \vec{g},  \vec{E}_\mathrm{sc} \right \rangle 
\right|^2\,.\label{eq:fiber_transmitted_power}
\end{eqnarray}
The first scalar product is determined by the mode
normalization. The second one, $\left \langle \vec{g},  \vec{E}_\mathrm{sc}
\right \rangle$, represents the projection of the scattered field onto the
collection mode. Modulo the normalization constant $\sqrt{P_\mathrm{in}}$, it
is identical to half the interference contribution in \Eref{Uf_int}, whose explicit expression was given in \Eref{Uf_int_result}. Therefore we find
\begin{equation}
1-\epsilon={P_\mathrm{out}\over P_\mathrm{in}}={1\over P_\mathrm{in}}\left|
   \sqrt{P_\mathrm{in}}-{P_\mathrm{sc}/2\over\sqrt{P_\mathrm{in}}}\right|^2 =
 \left|1-{R_\mathrm{sc}\over2}\right|^2\,. \label{eq:exact_extinction2}
\end{equation}
In the weak focusing regime where $R_\mathrm{sc}\ll1$, this translates again
into an extinction $\epsilon\approx R_\mathrm{sc}\,.$
For a focusing parameter $u=2.239$, we get a maximal extinction of
$\epsilon_\mathrm{max}=0.926$.
For the light scattered back into the excitation mode, we do not have to
consider the field $\vec{E}_\mathrm{F}$, and arrive similarly at
 $P_\mathrm{back}=\left| \left \langle \vec{g}, \vec{E}_\mathrm{sc} \right
   \rangle \right|^2=P_\mathrm{in}R_\mathrm{sc}^2/4$, or a reflectivity of
\begin{equation}
R=R_\mathrm{sc}^2/4\,.
\end{equation}

\section{Experiment}\label{sec:experiment}
\begin{figure}
  \begin{center}
    \includegraphics[width=1\columnwidth]{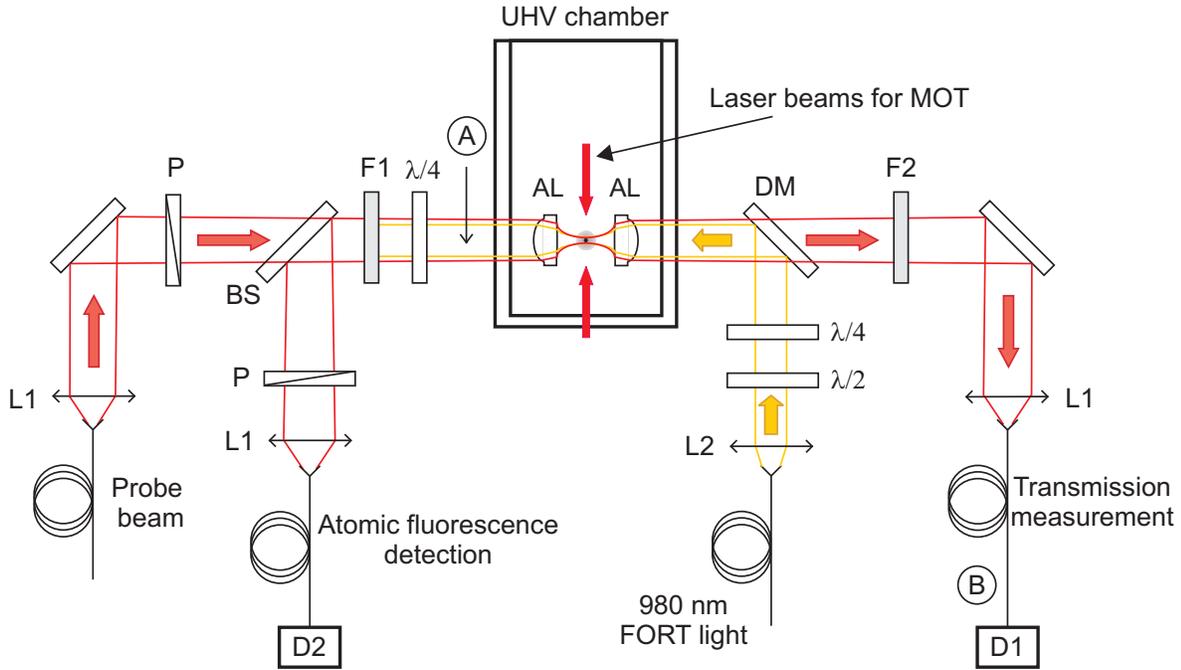}
    \caption{\label{experimental_setup}
      Experimental setup for measuring the extinction of a light beam by a
      single atom. AL: aspheric lens ($f=4.5\,$mm, full ${\rm NA}=0.55$), P:
      polarizer, DM: dichroic mirror, BS: beam splitter with 99\%
      reflectivity, $\lambda/4, \lambda/2$: quarter and half wave plates, F1:
      filters for blocking the 980\,nm FORT light, F2: interference filter
      centered at $780\,$nm, D1 and D2: Si-avalanche photodiodes. Four more
      laser beams forming the MOT lie in an orthogonal plane and are not
      shown.
    }
  \end{center}
\end{figure}

In this section, we consider the results of our experiment where we measured the
extinction of a Gaussian beam by a single $^{87}$Rb atom with different
focusing strengths, and compare the results to the above theoretical model.  
A detailed description of the experimental setup is reported
in~\cite{our_paper} and shown in Figure~\ref{experimental_setup}. 
Two aspheric lenses ($f=4.5$\,mm) are mounted in a UHV chamber
in a  confocal arrangement. A single $^{87}$Rb atom is localized in a
far-off resonant dipole trap (FORT) that is formed by 980\,nm light at the
focus of the lens pair. A probe beam is delivered from a
single-mode fiber and focused onto the atom by one lens, and picked up by the
other one. The confocal arrangement ensures that all of the incident probe
power is collected in the absence of an atom, thus implementing the scheme
discussed in the previous section. We use a circularly polarized probe to
optically pump the atom into a closed cycling transition.
After allowing some time for optical pumping, we measure the transmission of
the probe beam that is defined as the ratio of count rates at detector D1 when
the atom is present in the trap, to the count rate when the
atom is absent. Such a measurement is carried out for different probe
frequencies to obtain the transmission spectrum of a single Rb atom. The 
spectrum is fitted to a Lorentzian with the resonant frequency, the full width
at half maximum (FWHM) of the spectrum and its minimum value $T_\mathrm{min}$
on resonance as parameters. We obtained spectra for four different
input waists of the probe, thus measuring extinctions for
different focusing strengths. The observed FWHM never exceed 7.7 MHz, which is
close to the natural linewidth of the optical transition (6 MHz), so we
conclude that the atom  was successfully kept in a two-level system. For each
probe frequency, the probe power was adjusted such that the atom scatters
$\approx$ 2500 photons per second, way below saturation. The properties of
various transmission spectra obtained with different probe incident waists are
summarized in Table~\ref{table:results_summary}.
\begin{table}[h!b!p!]
\caption{Summary of transmission spectra of the probe for different focusing
  strengths $u$. Listed are $w_\mathrm{L}$: incident waist of the probe;
  $w_\mathrm{f}$ and $w_\mathrm{D}$: estimated paraxial focal waists of the
  probe beam and FORT, respectively; $\epsilon$ and W: maximum observed
  extinction value and FWHM of the transmission spectrum; $R_\mathrm{sc}$:
  scattering ratio for the focusing parameter; $\epsilon_\mathrm{theo}$:
  expected extinction according to \Eref{eq:exact_extinction2}.}
\label{table:results_summary}
\smallskip
\begin{indented}
\item[]\begin{tabular}{@{}llllllll}
\br
$w_\mathrm{L}$(mm) &$u$& $w_\mathrm{f}$($\mu$m) & $w_\mathrm{D}(\mu$m) & $\epsilon_\mathrm{max}$ (\%) & W (MHz) &$R_\mathrm{sc}$&$\epsilon_\mathrm{theo}(\%)$\\
\mr
0.5 &0.11& 2.23 & 2.0 & 2.38 $\pm$ 0.03 & 7.1 $\pm$ 0.2 &0.0362& 3.58\\
1.1 &0.24& 1.01 & 2.0 & 7.2 $\pm$ 0.1 & 7.4 $\pm$ 0.2 &0.1606& 15.41\\
1.3 &0.29& 0.86 & 1.4 & 9.8 $\pm$ 0.2 & 7.5 $\pm$ 0.2 &0.2157& 20.40\\
1.4 &0.31& 0.80 & 1.4 & 10.4 $\pm$ 0.1 & 7.7 $\pm$ 0.2 &0.2449& 22.99\\
\br
\end{tabular}
\end{indented}
\end{table}

We also carefully characterized the losses of the probe beam in its optical
path to ensure that our measured extinctions are not exaggerated by interference artefacts that can happen when certain elements in the transmission path preferentially filter more probe than the scattered light \cite{our_paper}.
From point A to point B in Figure~\ref{experimental_setup} we measured
53--60\% transmission without the atom in the trap. The losses are mostly
determined by 21\% loss through the uncoated UHV chamber walls and 17--24\% loss due
to the coupling into the single-mode fiber at the transmission measurement
channel. The coupling loss into the fiber increases as the input waist
of the probe beam $w_\mathrm{L}$ increases. Almost all losses can be ascribed to reflections at optical surfaces,
except for a 9--16\% re-coupling loss into a single mode fiber  due to
mode mismatch. We are thus reasonably confident that our measurement is free
from artefacts that may arise due to incomplete collection of the
probe.
\begin{figure}
  \begin{center}
    \includegraphics[width=1\columnwidth]{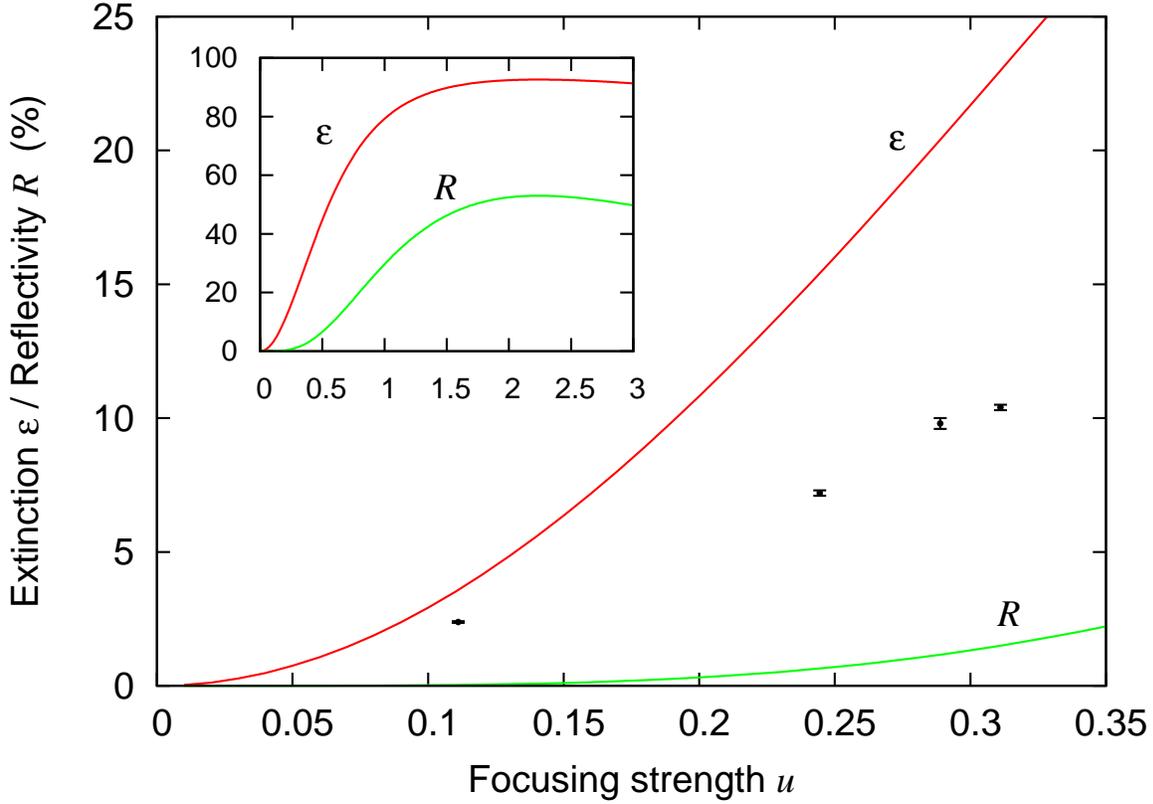}
    \caption{Experimentally measured extinction (black symbols) for several
      focusing strengths $u$ and predicted values for extinction $\epsilon$ and
      Reflectivity $R$ for a coupling into single mode optical fibers. The
      inset shows the prediction for much stronger focusing parameters.
      \label{fig:extinction_th+expt}}
    \end{center}
\end{figure}

In Figure~\ref{fig:extinction_th+expt} we compare the extinctions
obtained from the experiment with values predicted by
\Eref{eq:exact_extinction2}. Since the probe is recoupled into a single mode
fiber for every experimental point using a different lens that matches the probe
waist, and the coupling lens has a NA=0.55 (corresponding to $v=0.66$),
we can safely neglect any clipping, and use \Eref{eq:exact_extinction2} for
estimating the extinction.

Obviously the measured extinctions are smaller than those predicted,
especially for a larger focusing strength.  We see a few possible
reasons for this discrepancy. Firstly, the lenses we used in the experiment may
not be suffciently close to an ideal lens, since they were designed for a
situation with an additional window in the focusing part, which we did not
have in our experiment. Secondly, the interaction strength is significantly
affected by the motion of the atom in the dipole trap. While we do not have an
independent measure of the position fluctuation of the atom in the
trap, measurements in a similar trap showed a temperature of the atom in the
trap of around 100\,$\mu$K due to the overlapping MOT  \cite{weber:2006},
slightly below the Doppler temperature of 143\,$\mu$K of Rubidium.
With our trap frequencies of $\nu_\rho\approx70$\,kHz in transverse and
$\nu_z\approx20$\,kHz in longitudinal direction, this results in position
uncertainties of $\sigma_\rho\approx220$\,nm and $\sigma_z\approx780$\,nm,
respectively. The scattering 
ratio $R_\mathrm{sc}$ gets reduced due to the presence of the atom in regions
with a reduced excitation field, and  due to the
spatially dependent detuning in the optical dipole trap. In paraxial
approximation, we find an approximate reduction of the scattering rate due to
the lower average field of
\begin{equation}
  R_\mathrm{sc}'\approx
  R_\mathrm{sc}\left(1-2\sigma_\rho^2/w_\mathrm{f}^2\right)^2 
  \left(1-\sigma_z^2\lambda^2/(\pi^2w_\mathrm{f}^4)\right)\,,
\end{equation}
which results in a reduction of 2\% for $w_L=0.5$\,mm to 23\% for
$w_L=1.4$\,mm. The reduction in $R_\mathrm{sc}$ and therefore in $\epsilon$ in
our regime is about proportional to the temperature, so a
doubling of the temperature alone would explain the discrepancy between theory
and experiment already. On the other hand, the contribution due to a spatial
variation of the resonance frequency is less than 1\%.
Additionally, the presence of the atom away from the focal point reduces the
efficiency of the optical pumping.
However, the observed extinction ratios still exceed the prediction by the
parabolic wave front model in \cite{van_Enk:2001}.

The inset of Figure \ref{fig:extinction_th+expt} extrapolates the extinction
we could expect for much stronger focusing; as mentioned earlier, the maximal
extinction should reach 92.6\%; whether such lenses can be manufactured,
however, remains an open question.
We also depicted the reflectivity of the fiber-atom-fiber system, which should
reach values strong enough to be detected in an experimental setup. Extinction
and reflectivity don't match in this configuration, which means that there is
still a significant amount of light which is neither transmitted nor reflected
back into the optical fiber. Such losses are unavoidable for the described
coupling scheme, which still places the fiber-atom-fiber system in
disadvantage to a atom-cavity system in terms of success probability of
scattering into known modes.
If it would be possible to achieve a better
overlap of photonic modes with the dipole transition, such losses should 
be reduced.

\section{Conclusion}
We have demonstrated both theoretically and experimentally that a
substantial coupling efficiency of a light beam to a single atom can be
achieved by focusing a light beam with a lens. By modifying the model
given in \cite{van_Enk:2001}, we have constructed a focusing field compatible
with Maxwell equations that is suitable for the strong focusing
regime. High values for the extinction of light (up to 92\%) by a
two-level atom stationary at the focus under the assumptions of weak
on-resonant coherent probe are predicted. Within the limitations of our
current trap, our experimental results confirm the possibility of observing a
substantial extinction already for relatively weak focusing. 

The measured extinction depends on the particular collection configuration
of an experimental setup. It is thus not a fixed quantity for a given incident
field. As such, the scattering ratio as defined by \Eref{scatt_ratio} is a
better quantity to characterize the interaction strength between a weak
coherent field and an atom in free space, even though it loses a simple physical interpretation in the strong focusing regime. These results may also be of interest for experiments with single molecules \cite{molecule1,molecule2} and quantum dots \cite{quantumdots}.

\section*{Acknowledgement}
This work was partly supported by the National Research Foundation, the
Ministry of Education, Singapore, and by FRC grant R-144-000-174-112. The
authors would like to thank Gert Zumofen, Ilja Gerhardt, and Mark Dennis for
helpful discussions. ZC acknowledges financial support from ASTAR Singapore.

\appendix
\section{Power scattered by the atom in a coherent light field
  \label{App1}}

We briefly repeat the results for light scattered by a two-level atom
following \cite{cohen-tann:1992}. 
The steady-state population $\rho_{22}$ of the excited state in a two-level atom
exposed to a monochromatic field can be obtained from the optical Bloch
equations:
\begin{equation}
\rho_{22}=\frac{|\Omega|^2/4}{\delta^2+|\Omega|^2/2+\Gamma^2/4}
\label{population}
\end{equation}
Therein, $\Gamma$ is the radiative decay rate of the excited state,
\begin{equation}
\Gamma=\frac{\omega_{12}^3|d_{12}|^2}{3\pi\epsilon_0\hbar c^3}\,,
\label{decay}
\end{equation}
and $\Omega=E_\mathrm{A}|d_\mathrm{12}|/\hbar$ is the Rabi frequency for a
vanishing detuning $\delta=\omega-\omega_{12}$ of the driving field with
respect to the atomic transition frequency $\omega_{12}$. Therein,
$|d_\mathrm{12}|$ is the electrical dipole moment of the atom.

The optical power scattered by this atom is simply the product of energy
splitting, decay rate and population of the excited state: 
\begin{equation}
\label{scattered_power}
P_\mathrm{sc}=\rho_{22}\Gamma\hbar\omega_{12}
\end{equation}
For weak ($\Omega\ll\Gamma$) on resonant ($\delta=0$) excitation, the
scattered power becomes
\begin{equation}
\label{scatteredPower}
P_\mathrm{sc}=\frac{3\epsilon_0c\lambda^2E_\mathrm{A}^2}{4\pi}. 
\end{equation}
for an excitation field amplitude $E_A$ at the location of the atom.

\section{Transformation of local polarization by the lens \label{App2}}
To obtain the local polarization of the focusing field
in~\Eref{physical_focused_field}, we consider a point
P$(\rho,\phi,z)$ before the lens and an incident light field with polarization
\begin{equation}
\hat{\epsilon}_{in}=\hat{\epsilon}_+=\frac{\hat{x}+i\hat{y}}{\sqrt2}\,,
\end{equation}
or in the cylindrical basis,
\begin{equation}
\hat{\epsilon}_{in}=\frac{e^{i\phi}}{\sqrt2}\textrm{
}\hat{\rho}+\frac{ie^{i\phi}}{\sqrt2}\textrm{ }\hat{\phi}\,,
\end{equation}
where $\hat{\rho}=\cos{\phi}\textrm{ }\hat{x}+\sin{\phi}\textrm{ }\hat{y}$ and
$\hat{\phi}=-\sin{\phi}\textrm{ }\hat{x}+\cos{\phi}\textrm{ }\hat{y}$ are two
unit vectors along the radial and azimuthal directions respectively. 
The ideal lens leaves the azimuthal component unchanged but tilts the radial
component such that the local polarization of the field right after the lens
is perpendicular to the line FP in Figure~\ref{fig:geometry} (F is the focus
point), that is:
\begin{eqnarray}
\hat{\epsilon}_{F}&=&\left(\frac{\cos{\theta}\,e^{i\phi}}{\sqrt2}\,\hat{\rho}
  + \frac{\sin{\theta}\,e^{i\phi}}{\sqrt2}\,\hat{z}\right)
+ \frac{ie^{i\phi}}{\sqrt2}\,\hat{\phi}\\ 
&=&\frac{1+\cos\theta}{2}\,\hat{\epsilon}_+ + \frac{\sin\theta
  e^{i\phi}}{\sqrt{2}}\,\hat{z} +
\frac{\cos\theta-1}{2}e^{2i\phi}\,\hat{\epsilon}_-\,,\nonumber
\end{eqnarray}
where $\theta=\arctan(\rho/f)$ and
$\hat{\epsilon}_-=(\hat{x}-i\hat{y})/\sqrt2$.

\section{Decomposition of a field into modes with cylindrical symmetry\label{sec:cylmodes}}
For completeness, directly following \cite{van_Enk:2001}, we outline the main properties of the cylindrical modes $\vec{F_\nu}$, which form a complete orthogonal set to compose an electric field that satisfies the source-free Maxwell equations,
\begin{equation}
\label{field_expansion}
\vec{E}(t)=2\Re\left[\sum_\nu a_\nu\vec{F}_\nu e^{i\omega t}\right]\,,
\end{equation}
where the summation over $\nu$ is a short-hand notation for
\begin{equation}
\sum_\nu:=\int dk \int dk_z \sum_s \sum_m,
\end{equation}
and $a_\nu$ are arbitrary complex amplitudes. The modes are characterized by
four indices $\nu:=(k,k_z,m,s)$, where $k=\frac{2\pi}{\lambda}$ is the
wave vector modulus, $k_z=\vec{k}\cdot\hat{z}$ the wave vector component in
$z$-direction, $m$ an integer-valued angular momentum index, and $s=\pm1$ the
helicity. The dimensionless mode functions $\vec{F}_\nu$ in 
cylindrical coordinates $(\rho,z,\phi)$ given in \cite{vanenk:1994} are
\begin{eqnarray}
\fl \vec{F}_\nu (\rho,z,\phi)=&\frac{1}{4\pi}\frac{sk-k_z}{k}G(k,k_z,m+1)\hat{\epsilon}_- + \frac{1}{4\pi}\frac{sk+k_z}{k} G(k,k_z,m-1)\hat{\epsilon}_+ \nonumber\\
&- i\frac{\sqrt2}{4\pi}\frac{k_t}{k}G(k,k_z,m)\hat{z}\,,
\end{eqnarray}
where $k_t=\sqrt{k^2-k_z^2}$ is the transverse part of the wave vector,
$\hat{\epsilon}_\pm=(\hat{x}\pm i \hat{y})/\sqrt2$ are the two circular polarization vectors, and
\begin{equation}
G(k,k_z,m)=J_m(k_t \rho)e^{ik_z z}e^{im\phi},
\end{equation}
with $J_m$ the $m$-th order Bessel function. As we are interested in a
monochromatic beam with a fixed value of $k=2\pi/\lambda$ propagating
in the positive $z$ direction ($k_z>0$), the set of mode indices is reduced
to $\mu:=(k_t,m,s)$ where, for convenience, $k_t$ is taken as a mode
index instead of $k_z$. Now, we introduce the notation
\begin{equation}
\sum_\mu:= \int dk_t \sum_s \sum_m
\end{equation}
for a complete summation over all possible modes. For a fixed $k$ the modes
$\vec{F}_\mu$ are orthogonal with respect to the scalar product
\begin{equation}\label{orthogonalrelation1}
\left\langle\vec{F}_\mu,\vec{F}_{\mu'}\right\rangle = \int_S\textrm{
}\vec{F}_\mu^\ast (\vec{r})\cdot\vec{F}_{\mu '}(\vec{r})\,dS = 
\delta(k_t-k_t')\delta_{mm'}\delta_{ss'}/(2\pi k_t)\,,
\end{equation}
where $S$ is a plane perpendicular to the $z$ axis. This scalar product can
thus be used to find the amplitudes of the modes $\mu, \mu'$ in an arbitrary
electric field compatible with the Maxwell equations.



\section*{References}
\begin{thebibliography}{14}
\bibitem{cirac:1997} Cirac J~I, Zoller P, Kimble H~J and Mabuchi H 1997 {\it  \PRL} {\bf 78} 3221--27 
\bibitem{duan:2001} Duan L~M, Lukin M~D, Cirac J~I and Zoller P 2001 {\it  Nature} {\bf 414} 413--18
\bibitem{walls:1990} Savage S~M, Braunstein S~L and Walls D~F 1990 {\it Optics
    Letters} {\bf 15} 628--30
\bibitem{rosenfeld:07}Rosenfeld W, Berner S, Volz J, Weber M, and Weinfurter H
  2007 {\it \PRL} {\bf 98} 050504
\bibitem{leuchs:2007} Sondermann M, Maiwald R, Konermann H, Lindlein N,
  Peschel U and Leuchs G 2007 {\it Applied Physics B} {\bf 89} 489--92

\bibitem{van_Enk:2001} van Enk S~J and Kimble H~J 2001 {\it Physical Review A}
  {\bf 63} 023809
\bibitem{wineland:87}Wineland D~J, Itano W~M, Bergquist J~C 1987 {\it Optics
    Letters} {\bf 12}, 389
\bibitem{molecule1}Gerhardt I, Wrigge G, Bushev P, Zumofen G, Agio M, Pfab R,
  and Sandoghdar V 2007 {\PRL} {\bf 98}, 033601
\bibitem{our_paper} Tey M~K, Chen Z, Aljunid S~A, Chng B, Huber F, Maslennikov
  G and Kurtsiefer C 2008 {\it Nature Physics} {\bf 4} 924--27
\bibitem{cohen-tann:1992} Cohen-Tannoudji C, Grynberg G and Dupont-Roc J 1992
  {\it Atom-Photon Interactions: Basic Processes and Application} (Wiley, New
  York)

\bibitem{wolf:1959} Richards B and Wolf E 1959 {\it Proceedings of the Royal
    Society A} {\bf 253} 358--79
\bibitem{vanenk:1992} van Enk S~J and Nienhuis G 1992 {\it Optics
  Communications} {\bf 93} 147--58
\bibitem{beijersbergen:1993} Beijersbergen M~W, Allen L, van der Veen H~E~L~O
  and Woerdman J~P 1993 {\it Optics Communications} {\bf 96} 123--32
\bibitem{jackson:CE} Jackson J D 1975 {\it Classical Electrodynamics} (Wiley,
  New York), 2nd ed
\bibitem{born:PO} Born M and Wolf E 1975 {\it Principles of Optics} (Pergamon,
  Oxford)

\bibitem{bohren:1983} Bohren C 1983 {\it American Journal of Physics} {\bf
    51}, 323
\bibitem{bassett:1986} Bassett I~M 1986 {\it Journal of Modern Optics} {\bf
  33} 279--86
\bibitem{paul:1983} Paul H and Fisher R 1983 {\it Soviet Physics Uspekhi} {\bf 26}, 923
\bibitem{davis:2001} Davis R C and Williams C C 2001 {\it Journal of the
    Optical Society of America} {\bf 18} 1543
\bibitem{zumofen:2008} Zumofen G, Mojarad N M, Sandoghdar V, and Agio M 2008 {\PRL} {\bf 101} 180404

\bibitem{bohren:ASLSP} C. F. Bohren and D. R. Huffman 1983 {\it Absorption and
Scattering of Light by Small Particles} (Wiley, New York).
\bibitem{sheppard:1997} Sheppard C J R and T\"{o}r\"{o}k P 1997 {\it Journal of Modern Optics} {\bf 44} 803
\bibitem{mojarad:2008} Mojarad N M, Sandoghdar V, and Agio M 2008 {\it Journal
    of the Optical Society of America B} {\bf 25} 651
\bibitem{weber:2006} Weber M, Volz J, Saucke K, Kurtsiefer C and
  Weinfurter H 2006, {\it Physical Review A} {\bf 73} 043406
\bibitem{molecule2} Wrigge G, Gerhardt I, Hwang J, Zumofen G and Sandoghdar V
  2008 {\it Nature Physics} {\bf 4} 60--66 

\bibitem{quantumdots}Vamivakas A et al. 2007 {\it Nano Letters} {\bf 7} 2892--96
\bibitem{vanenk:1994} van Enk S~J and Nienhuis G 1994 {\it Journal of Modern
  Optics} {\bf 41} 963--77

\end{thebibliography}
\end{document}